# DAΦNE EXPERIENCE WITH NEGATIVE MOMENTUM COMPACTION

M. Zobov, D. Alesini, M.E. Biagini, A. Drago, A. Gallo, C. Milardi, P. Raimondi,
B. Spataro, A. Stella, LNF-INFN, Frascati, Italy

*Abstract*

There are several potential advantages for a collider operation with a lattice with negative momentum compaction factor (alfa, $\alpha_c$). Since the lattice of the Frascati e+e- Φ-factory DAΦNE is flexible enough to provide collider operation even with alfa < 0, we have exploited this possibility for an experimental study of the beam dynamics. The negative momentum compaction lattices have been successfully implemented and stable 1 A currents have been stored in both electron and positron rings without any problem for RF cavities and feedback systems operation. First collisions have been tested at low currents. In this paper we describe the experimental results and compare them with expectations and numerical simulations. Present limitations to DAΦNE operation with alfa < 0 and ways to overcome them are also discussed.

## INTRODUCTION

The e+e- collider DAΦNE, the 1.02 GeV c.m. Frascati Φ-factory [1], has reached a peak luminosity of about $1.5 \times 10^{32}$ cm$^{-2}$s$^{-1}$ and a daily integrated luminosity of 10 pb$^{-1}$ [2]. At present the DAΦNE Team is discussing several scenarios for future upgrades [3] and different ideas aimed at increasing the luminosity are being studied theoretically and tested experimentally. One of such ideas which does not require additional costs since it relies only on the flexibility of the lattice is to realize an optical structure with a negative momentum compaction factor.

There are several potential advantages for beam dynamics and luminosity performance of a collider with a negative momentum compaction factor:

- The bunch length is shorter since the wake potential is focusing. A shorter bunch is preferable for both peak luminosity and beam-beam lifetime improvement since the transverse beta functions at the interaction point (IP) can be reduced proportionally to the bunch length. Moreover, the Piwinski angle becomes lower in collisions with a crossing angle, as in the case of DAΦNE.
- Qualitative considerations confirmed by numerical simulations have shown that a lattice with negative momentum compaction can make the longitudinal beam-beam effects less harmful [4], improving beam lifetime and decreasing transverse beam size blow up. Moreover, $\alpha_c < 0$ avoids coherent and incoherent instabilities arising when the longitudinal beam-beam kick gets comparable with the RF voltage [5].
- It has been shown that for certain vacuum chamber coupling impedances the microwave instability threshold can be higher with negative alfa [6]. This, however, has not been confirmed in few dedicated experiments [7, 8, and 9] and must be checked for the DAΦNE impedance.
- The single bunch head-tail instability with negative momentum compaction takes place with positive chromaticity. This means that one can operate a storage ring with a 'natural' negative chromaticity without correcting sextupoles.

Thus, a dedicated machine experiment aimed at studying DAΦNE performance with $\alpha_c < 0$ was scheduled and then performed with the following principal goals:

- to check the reliability of the lattice model by varying the momentum compaction in a wide range,
- to prove bunch shortening with the wake fields in DAΦNE and to study the microwave instability,
- to study RF cavities and feedbacks performance and to investigate the high current multibunch dynamics,
- to perform beam-beam collisions with negative alfa.

This paper describes the obtained results and analyzes the experimental data comparing them with expectations and numerical simulations. Present limitations to DAΦNE operation with negative momentum compaction and possible ways to overcome them are also discussed.

## LATTICE MODIFICATION

DAΦNE is a double ring collider sharing two interaction regions; each ring consists of a long external arc and a short internal one based on a quasi-achromatic cell built using a bending-wiggler-bending sequence. The presence of the wiggler in the region of maximum dispersion doubles the radiation emitted in the dipoles reducing damping times. Moreover the optics flexibility (all quadrupoles are independently powered) makes it possible to tune the beam emittance by varying the dispersion in the wiggler at constant field.

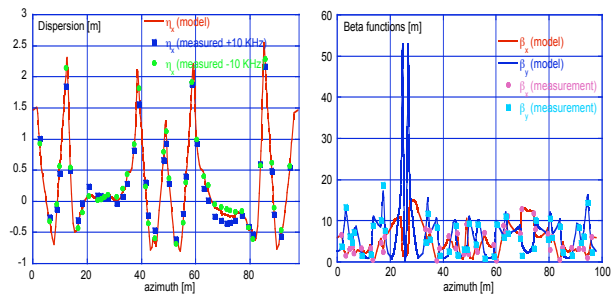

Figure 1: Optical functions of the positron ring with $\alpha_c$=-0.019 (solid lines: modeling; dots: measurements).

The behaviour of the dispersion function in the arcs determines also the value of the momentum compaction factor $\alpha_c$. In the nominal working conditions $\alpha_c$ = 0.02-0.024, but this parameter can be easily moved toward negative values by making the dispersion function negative alternatively at the entrance and at the exit of the arcs along the beam direction. The dispersion adjustments should preserve the transverse betatron functions and the relative betatron phase advances in the two interaction regions and minimize their variation along the rings. This requires a general reconfiguration of the currents in all quadrupoles. During a short period devoted to machine studies two different value of $\alpha_c$ have been investigated: $\alpha_c = -.02$ (in both rings) and $\alpha_c = -.036$ (only in the electron ring). After applying the new optics with negative momentum compaction and slight lattice optimization the ring has been completely characterized by measuring betatron functions, first and second order dispersion, chromaticity with and without sextupoles and the synchrotron frequency before starting systematic bunch length measurements. The optics implementation and measurements required less than one day per ring. In both rings the experimental data are in a good agreement with the optics model both for the linear and non-linear parts [10]. As an example, the transverse betatron functions and the dispersion measured in the positron ring are compared with the computed ones in Fig. 1.

## BEAM DYNAMICS

Many beam dynamics experiments, including bunch length measurements, study of high current multibunch operation and first beam-beam collisions, were performed with $\alpha_c = -0.019$ in the positron ring and $\alpha_c = -0.021$ in the electron one. The absolute values of the negative $\alpha_c$ were chosen to be as close as possible to the positive one in the normal collider operating conditions.

### Bunch Shortening

Impedances and wake fields are well known for both the positron [11] and the electron [12] ring. Numerical simulations based on the wake fields reproduce well bunch length, charge distribution inside the bunch and the microwave instability threshold with positive momentum compaction. These wakes have been also used to simulate bunch shortening with negative alfa [13]. According to the predictions the bunch length should be shorter by 50÷70% than in nominal operating conditions.

In order to confirm these predictions the bunch length in both rings has been measured by means of a streak camera. The measurement set up is discussed elsewhere [14]. The results are shown in Fig. 2 comparing bunch length for positive and negative $\alpha_c$. It can be observed that the bunch is substantially shorter with $\alpha_c < 0$, it shortens up to the microwave threshold and then starts growing. As predicted by simulations, for the positron ring the microwave instability threshold is only slightly lower than with the positive alfa: it is shifted from 9÷10 mA down to 7 mA. No instability sidebands have been observed up to 16 mA/bunch which is higher than in the present working conditions. No intensity loss due to the head-tail instability has been detected for positron bunch currents as high as 40 mA with high negative chromaticities.

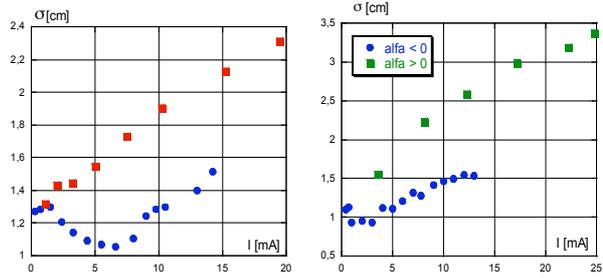

Figure 2: Bunch length as a function of bunch current in positron (left) and electron (right) rings (squares: positive alfa; circles: negative alfa).

In the electron ring the bunch length is also shorter with the negative momentum compaction, but the microwave instability threshold is significantly lower than in the positron ring. This is explained by the larger coupling impedance of the electron ring (by about a factor 2) [12].

### Multibunch Operation

The RF cavity operation with negative momentum compaction factor required only:
- change of the RF phase by about 170 degree (1.3 ns in time units) with respect to the injection complex, since the synchronous phase in this case is on the positive slope of the RF voltage;
- re-adjustment of the tuning loop to get a certain positive detuning of the accelerating mode frequency, i.e. to provide some Robinson damping already at low current.

High current operation in both rings has been tested by injecting 100 consecutive bunches (out of a maximum of 120). No particular problem has been detected in managing the multibunch dynamics. Surprisingly, feedbacks were not necessary to store 400 mA in the electron ring. In order to obtain about 1 A of stable currents in both electron and positron beams the following procedures and set ups of the feedback systems have been applied:
- Longitudinal feedback: 'standard' FIR filter (broadband, 'low gain'), front-end and back-end retiming in both rings.
- Vertical feedback: only back-end and front-end retiming in both rings.
- Horizontal feedback: back-end and front-end retiming only in the positron ring; no horizontal motion in the electron ring till about 1A.

The maximum storable currents at the level of 1 A were limited by injection saturation. Further current increase would require careful adjustment of injection closed bumps and dynamic aperture optimization.

### First Beam-Beam Collisions

Optimization of beam-beam collisions has been carried out at low beam currents (less than 100 mA/beam) by measuring the luminosity as a function of calibrated vertical and horizontal bumps at the interaction point (IP). The average geometric rms beam sizes ($\Sigma_{x,y}$) at IP, extracted from these measurements, have been minimized by adjusting collider parameters such as tunes, coupling, beams overlap etc. After tuning we managed to reduce the vertical capital sigma $\Sigma_y$ down to 8.2 μm and the horizontal one to 1 mm. These $\Sigma_{x,y}$ are comparable with the best values obtained in DAΦNE with positive momentum compaction.

By increasing the beam currents in collision the best obtained luminosity was $2.5 \times 10^{31}$ cm$^{-2}$s$^{-1}$ while colliding 300 mA electrons against 300 mA positrons. This result corresponds to a specific luminosity of $2.5 \times 10^{28}$ cm$^{-2}$s$^{-1}$mA$^{-2}$ defined as the single bunch luminosity divided by the product of colliding bunch currents. One can compare this number with the typical specific luminosity in high current collisions in DAΦNE with $\alpha_c > 0$ which ranges between 1 and 1.5 in units of $10^{28}$ cm$^{-2}$s$^{-1}$mA$^{-2}$. However, such comparison is not conclusive since so far with $\alpha_c < 0$ we did not collide as high currents as they are in routine DAΦNE operation with $\alpha_c > 0$.

## LIMITATIONS AND CURES

The principal limitation that forced us to suspend high current beam-beam collisions was a very fast growth of the electron vertical beam size above a threshold of 3-4 mA per bunch. A strong correlation between the longitudinal microwave instability and the vertical size blow up has been found. As can be seen in Fig.3, the single bunch longitudinal instability threshold at about 3 mA (a) corresponds to a vertical size blow up threshold at 300 mA (b) for an electron beam of 90 bunches.

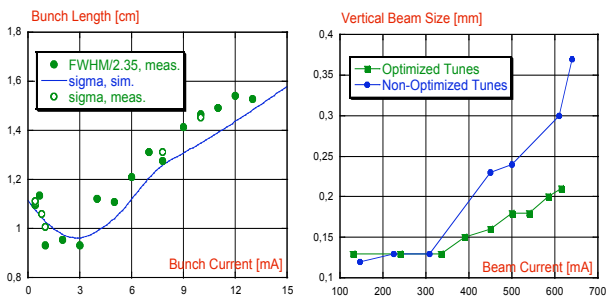

Figure 3: a) Bunch length as a function of bunch current (solid line: simulations; dots: measurements); b) Vertical beam size as a function of beam current (90 bunches) without (blue) and after tune optimization.

Such a correlation is not typical only for operation with $\alpha_c < 0$, but it has been also detected and studied for the DAΦNE routine lattice with $\alpha_c > 0$ [15]. However, since the microwave threshold is lower for negative alfa, the effect is much more pronounced. It is particularly strong for the electron ring which has a higher impedance.

The only way to overcome this limitation is to move the microwave threshold beyond the nominal bunch current.

Two ways are being followed to increase the threshold. First, during the current DAΦNE shutdown 2 m long ion clearing electrodes in the wiggler sections, which account for almost half of the electron ring impedance budget, will be removed. Second, according to the Boussard criterion [16], the threshold increases rapidly with the absolute value of the momentum compaction ($I_{th}$ scales as $\alpha_c^{3/2}$). Exploiting this property, one day before the shutdown a lattice with $\alpha_c = -0.036$ has been implemented (instead of $-0.021$) with encouraging results [17]: the threshold has been shifted from 3 up to 7÷8 mA/bunch while the bunch length at the nominal bunch current remained by 50% shorter than in the operational lattice with $\alpha_c > 0$. We plan to resume studies of the negative momentum compaction beam dynamics after the shut down.

## SUMMARY

- The DAΦNE optics model has proved to be reliable in providing collider operation with the momentum compaction factors in the range from +0.034 [10] to -0.036.
- With negative alfa bunches in both rings shorten as predicted by numerical simulations. It has been possible to store high bunch currents with large negative chromaticities.
- No hard limit has been seen in multibunch operation. About 1 A stable currents have been stored in both rings.
- At beam currents up to 300 mA/beam a good specific luminosity has been obtained in beam-beam collisions.
- Higher current collisions have been prevented by fast growth of the electron beam vertical size with current (single beam effect). We hope to overcome this limitation by reducing the coupling impedance in the electron ring and/or applying an optics with higher $|\alpha_c|$.

## REFERENCES

[1] G. Vignola, PAC93, p. 1993.
[2] A. Gallo et al., this conference, MOPLS028.
[3] M. Zobov et al., PAC05, p. 112.
[4] V. V. Danilov et al., HEACC'92, p. 1109.
[5] V. V. Danilov et al., PAC91, p. 526.
[6] S. X. Fang et al., PAC95, p. 3064.
[7] A. Nadji et al., EPAC96, p.676.
[8] M. Hosaka et al., Nucl.Instr.Meth. A407, 234(1998).
[9] H. Ikeda et al., e-print Archive:physics/0401155.
[10] www.lnf.infn.it/acceleratori/dafne/report/KLOE_acn_Feb_05.pdf
[11] M. Zobov, e-print Archive:physics/0312072.
[12] B. Spataro and M. Zobov, DAΦNE Note G-64, 2005.
[13] www.lnf.infn.it/conference/d2/TALKS/zobov1.pdf
[14] A. Ghigo et al., EPAC2002, p. 1494.
[15] www.lnf.infn.it/acceleratori/dafne/report/High_momentumCompaction.pdf
[16] D. Boussard, CERN LABII/RF/INT/75-2, 1975.
[17] www.lnf.infn.it/acceleratori/dafne/report/Zobov_Neg_Alfa_036_Mar06.pdf